\documentstyle[11pt]{article}

\input epsf
\title{\bf Black Holes and Causal Structure in Anti-de Sitter Isometric Spacetimes}
\author{S\"oren Holst\thanks{Department of Physics,
    Stockholm University, Box 6730, S-113 85 Stockholm, Sweden.\newline
    e-mail: holst@vanosf.physto.se}
  \and Peter Peld\'{a}n\thanks{Department of Physics,
    Stockholm University, Box 6730, S-113 85 Stockholm, Sweden.\newline
    e-mail: peldan@vanosf.physto.se}}
\date{}
\begin{document}
\maketitle
\begin{abstract}
  The observation that the 2+1 dimensional BTZ black hole can be
  obtained as a quotient space of anti-de Sitter space leads one to
  ask what causal behaviour other such quotient spaces can display. In
  this paper we answer this question in 2+1 and 3+1 dimensions when
  the identification group has one generator. Among other things we
  find that there does not exist any 3+1 generalization of the {\em
    rotating} BTZ hole. However, the non-rotating generalization
  exists and exhibits some unexpected properties. For example, it
  turns out to be non-static and to possess a non-trivial apparent
  horizon.
\end{abstract}
\section{Introduction}
In 1992 Ba\~nados {\it et al} \cite{Banados} reported a new 2+1
dimensional solution to Einsteins equations in vacuum with a negative
cosmological constant. As all such solutions this one was locally
isometric to anti-de Sitter space (adS-space)---the negatively curved
analogue of Minkowski space---and could even be constructed from it
simply by means of an identification of points \cite{BH}. The
surprising fact with this solution, so easily obtained, was that it
described a black hole---a very ``minimal'' one, containing just the
necessary ingredients in order to fulfil the defining properties.
Hence this solution gave rise to a rather large amount of papers (see
Carlip \cite{Carlip} and Mann \cite{Mann} and references therein).

However, not much has been said about its 3+1 dimensional analogues.
In ref. \cite{Haal} some examples of such generalizations were given,
and ref. \cite{BanadosMultiDim} discusses adS black holes in even higher
dimensions, especially the 4+1 case. Here we intend to give a more
systematic treatment of the subject, following the method of appendix
A in ref. \cite{BH} to classify the different spacetimes that can be
obtained by making identifications in 3+1 dimensional adS-space using
a one dimensional subgroup of its symmetry group SO(2,3). We
will see that the different cases will exhibit a great variety of
causal structures, most of which, unfortunately, involve naked regions
with closed timelike curves. In particular, we will show that there
does not exist any 3+1 dimensional generalization of the {\it
  rotating} BTZ black hole. It is only the non-rotating special case
that has a 3+1 dimensional counterpart.

The route in this paper will be as follows. First we will explain how
to classify all spacetimes obtainable as a quotient of adS-space by
one of its one generator subgroups. This
amounts to classifying the elements in SO(2,3), or equivalently, the
Killing fields of 3+1 adS-space. Given a particular Killing field we
will see how to deduce the causal structure of the corresponding
(identified) spacetime. Then we will perform the classification of
SO(2,3), and compare the result with the classification of SO(2,2),
performed in ref. \cite{BH}. A particular choice of coordinates will
enable us to visualize our conclusions concerning the different causal
structures. At the end we will take a closer look at the only 3+1
dimensional black hole that can be obtained by the methods used in
this paper---the 3+1 dimensional analogue of the non-rotating BTZ
black hole. That solution actually turns out to be non-static, and to
contain an apparent horizon which does {\it not} coincide with the
event horizon.

\section{Causal structure and symmetries}
As mentioned above, the BTZ black hole can be obtained by an
identification of points in adS-space. More specificially, one
identifies all points that are mapped into each other under a
particular discrete symmetry $\Gamma$ of adS-space. In other words,
the BTZ-solution could be considered as the quotient space
$[adS]/{\cal G}_{_\Gamma}$, where ${\cal G}_{_\Gamma}$ is the
group generated by $\Gamma$: ${\cal G}_{_\Gamma} = \{
\Gamma^n ;n \in \bf Z \}$. Furthermore, this $\Gamma$ is obtainable by
exponentiating a local symmetry of SO(2,2)---the symmetry group of 2+1
adS-space---that is, $\Gamma = e^{\alpha \xi}$ where $\xi$ is a Killing
field and $\alpha$ is a finite parameter fixing the ``size'' of
$\Gamma$.

So by simply taking the quotient space of adS-space with a particular
$\Gamma$ we can obtain such interesting objects as black holes. The
question then arises, what other causal structures could be obtained
by choosing $\Gamma$ differently? The goal of this paper is to answer
this question in 2+1 and in 3+1 dimensions.

There is one point about the notation that should be made clear from
the beginning: If ${\cal G}_{_\Gamma}$ does not act properly
discontinuously, which for example may be the case if the generating
Killing field contains rotation, then, strictly speaking, we cannot
take the quotient $[adS]/{\cal G}_{_\Gamma}$ without first removing
points from adS-space, and maybe extend it to some covering space. In
what follows we will ignore such technicalities and just write
$[adS]/{\cal G}_{_\Gamma}$ even if ${\cal G}_{_\Gamma}$ fails to act
properly discontinuously. In each case the meaning should be clear
anyway.

We must also clarify what we mean by ``different $\Gamma$:s''. Of
course, we are not interested in comparing symmetries that differ only
by a global isometry, because they have to yield the same spacetime
when used for identification. Thus, if two generators $\Gamma_1$ and
$\Gamma_2$ satisfy
\begin{equation}
  \label{similarity}
  \Gamma_2 = T^{-1} \Gamma_1 T \hspace{15mm} T \in {\rm SO}(2,d-1)
\end{equation}
where $d$ is the spacetime dimension (3 or 4 here), they will be
considered to correspond to the same symmetry. Our task is then to
classify SO(2,\,$d-1$), or equivalently the Killing fields for
adS-space, up to this equivalence relation. For $d=3$ this was done in
ref. \cite{BH} and in the next section we will compare that result
with the corresponding one for $d=4$.

For now, consider this to be done. The next question then is how we,
given a Killing field $\xi$, can deduce the causal structure of
$[adS]/{\cal G}_{_\Gamma}$ where ${\cal G}_{_\Gamma}$ is generated by
$\Gamma = e^{\alpha \xi}$ for some fixed $\alpha$. First, obviously
enough, if $\xi$ is timelike in some region in adS-space, then, when
we identify points connected by $e^{\alpha \xi}$, closed timelike
curves will result in this region, consisting simply of the (closed)
Killing field lines there.\footnote{True adS-space has a cyclic time
  and hence contains closed timelike curves in itself. In this paper
  we will actually be working with its universal covering, see
  equations (\ref{sausagecoord}) and the following comments.} In
general, there could be closed timelike curves extending outside this
region too, but at least a small segment of them would have to be
inside this region.  Therefore, an observer not entering the region
where $\xi$ is timelike, that is, not passing the hypersurface
$\xi^\mu \xi_\mu = 0$, would not be able to travel on a closed
timelike curve. Thus, to clarify the causal structure of
$[adS]/{\cal G}_{_\Gamma}$ we would like to know how different observers,
in the region where $\xi$ is spacelike, are related to the surface
where $\xi$ is null. For this reason, from now on we adopt the
following definitions for the quotient space $[adS]/\{e^{n\alpha\xi}; n
  \in {\bf Z} \}$ for some Killing field $\xi$ and a fixed $\alpha$:

First, the hypersurface $\xi^\mu \xi_\mu = 0$ will be called the {\it
  singularity}. This definition is consistent with the language used
in ref. \cite{Banados} and \cite{BH}. In what follows we will be
interested in the causal structure with respect to this singularity.
This of course actually means, the causal structure with respect to
the region where timelike curves may close, something that the reader
who feels uncomfortable with this terminology may wish to keep
in mind.

The horizons are then defined as usual. The {\it event horizon} is the
boundary of the past of the future null infinity, $\dot{J}^-({\cal I}^+)$,
where the word ``infinity'' actually refers to the part of infinity
where $\xi$ is spacelike, that is, the part relevant to an observer
not allowed to pass the singularity. When an observer passes this
horizon he loses his chances to reach infinity. If there are
several disconnected infinities there is one event horizon for each of
them.

The {\it inner horizon} is the boundary to the future of the singularity,
$\dot{J}^+(\{\xi^\mu \xi_\mu = 0\})$. An observer passing this horizon
will be able to see the singularity, without falling into it.

Note that to apply these definitions, that is, to find the causal
structure in $[adS]/{\cal G}_{_\Gamma}$, the only thing we need is the
behaviour of the generating Killing field $\xi$ in adS-space. The parameter
$\alpha$ will not affect the causality and we do not have to obtain an
explicit metric for the quotient space, or know precisely which points
that are identified with each other. We do not even have to know very
much about the Killing field itself, only the properties of the
hypersurface $\xi^\mu \xi_\mu = 0$, especially near infinity.

Let us see what we can say about this singularity hypersurface in
general. Consider the scalar field defined by $f(x) = \xi^\mu
\xi_\mu$, where $\xi$ is some Killing field. The normal to hypersurfaces
where $f(x)$ is constant is
\[
\nabla_\rho f(x) = \nabla_\rho \xi^\mu \xi_\mu = 2 \xi^\mu \nabla_\rho
\xi_\mu
\]
which implies that
\[
\xi^\rho \nabla_\rho f(x) = 2 \xi^\rho \xi^\mu \nabla_\rho \xi_\mu = 0
\]
since $\xi$ is a Killing field: $\nabla_{(\rho} \xi_{\mu)} = 0$. Thus
$\xi$ is orthogonal to the normal of the hypersurfaces with $f(x)$
constant, and hence lies in such surfaces. In particular, the
Killing field is tangent to the singularity $f(x) = 0$. More
intuitively, this has to be the case, since otherwise the singularity
would not be mapped into itself under the isometry corresponding to
$\xi$.

From this we see that except where $\xi^\mu = 0$, the singularity
surface has to be either timelike or lightlike, since it has a tangent
that is null. If it is timelike it necessarily is naked with respect
to some region, that is, visible for some observers in the spacetime.

For the 2+1 dimensional case we can draw even more restrictive
conclusions about the singularity. Later, in order to visualize our
results, we will use a conformal picture of adS-space, that is,
coordinates where points infinitely far away from the origin are
mapped into a finite distance, and thus, where infinity itself is
represented as a hypersurface. The Killing field is well-defined in
this hypersurface by a limit procedure, and if it is non-zero there it
has to be tangent to infinity since otherwise adS-space would not
be mapped into itself by the corresponding symmetry. Now, if the
singularity and the infinity meet each other, they do so in a one
dimensional line, and since the Killing field is tangent to both of
them, this line has to be a Killing field line. Further, since the
field is null at the singularity, this line must be lightlike. (This,
in turn, is very restrictive for the shape of a possible event
horizon.)

In 3+1 dimensions we can not draw such a conclusion, because there the
singularity meets infinity in a two dimensional
surface, which then must have a null direction. This is not very
restrictive because this surface could still be either timelike or
lightlike.

\section{Classification of adS-isometries}
Up to now, what we have said is quite general. Not only did we avoid
specifying a particular symmetry or Killing field to be used for the
identification---we did not even use any properties of the playground
itself: Anti-de Sitter space. Actually, we have not even defined it
yet. The previous section holds for every quotient space of the form
$[{\cal M},g^{\mu\nu}]/\{\Gamma^n ; n \in {\bf Z} \}$ where $\Gamma$
is generated by a continuous symmetry $\xi$ on the spacetime
$[{\cal M},g^{\mu\nu}]$.

Now we will specialize to adS-space and discuss its isometries. This
will enable us, in the next section, to draw conclusions about the
different causal stuctures in all quotient spaces of the form
$[adS]/e^{\alpha\xi}$. We begin by defining 3+1 adS-space as the
hyperboloid
\begin{equation}
  \label{hyperboloid}
  X^2 + Y^2 + Z^2 - U^2 - V^2 = -1
\end{equation}
embedded in the flat 5 dimensional space with metric
\begin{equation}
  \label{5-metric}
  ds^2 = dX^2 + dY^2 + dZ^2 - dU^2 - dV^2
\end{equation}
The 2+1 dimensional case is defined by the same equations but with $Z
= 0$.

A natural set of ``base Killing vector fields'' are given by
\begin{equation}
  \label{baseKilling}
  J_{ab} = x_b \frac{\partial}{\partial x^a} - x_a \frac{\partial}{\partial x^b}
  \hspace{15mm} {\rm where} \;\; x^a = (U,V,X,Y,Z)
\end{equation}
in terms of which a general Killing field $\xi$ may be written as
\[
\xi = \frac{1}{2} \omega^{ab} J_{ab} = \omega^{ab} x_b \partial_a
\]
where $\omega^{ab} = - \omega^{ba}$. Hence, a general Killing field may
be characterized with the matrix $\omega^a_{\;b}$. In fact, since
\[
\xi x^c = \omega^a_{\;b} x^b \partial_a x^c = \omega^c_{\;b} x^b
\]
it is simply the generator of the infinitesimal symmetry
transformation $\delta x^a = \epsilon \xi x^a$:
\[
x'^a = x^a + \delta x^a = x^a + \epsilon \omega^a_{\;b} x^b =
(\delta^a_{\;b} + \epsilon \omega^a_{\;b}) x^b
\]

To classify the isometries of adS-space then amounts to classifying
the matrices $\omega^a_{\;b}$ up to the equivalence relation
(\ref{similarity}), that is,
\begin{equation}
  \label{omegasim}
  \omega'^{\;a}_{\;\;\;b} = (T^{-1})^{a}_{\;c} \omega^c_{\;d}
  T^d_{\;b}
\end{equation}
where $T^a_{\;b} \in {\rm SO}(2,d-1)$. However, we could equally well
say that we want to classify all antisymmetric matrices $\omega_{ab}$
up to similarity (that is, (\ref{omegasim}) with an arbitrary $T$),
since for such matrices (\ref{omegasim}) is fulfilled if and only if 
$T^a_{\;b} \in {\rm SO}(2,d-1)$.\footnote{Strictly speaking, if
  $\omega_{ab}$ and $\omega'_{ab}$ are anti-symmetric (\ref{omegasim})
  is fulfilled if and only if $T^a_{\;b} \in {\rm O}(2,d-1)$, that is,
  the determinant of $T$ could be $-1$. In practice this means that
  the classification below does not distinguish between two Killing fields
  that are the mirror image of each other, and they will belong to the
  same type in table~1 or 2 below.} 

Now, any diagonalizable matrix may be uniquely characterized, up to
similarity, by its eigenvalues. Hence, if every $\omega^a_{\;b}$ were
diagonalizable, which they of course are not, they could be classified
according to their eigenvalues. In our case we have to use a slightly
more sophisticated method than that. Recall that {\it any} matrix $M$
may be written as $PNP^{-1}$ were $N$ is not necessarily diagonal, but
at least on Jordan's normal form, that is (for 5 dimensions)
\[
N = \left( 
      \begin{array}{ccccc}
        \lambda_1 & a_1 & 0 & 0 & 0 \\
        0 & \lambda_2 & a_2 & 0 & 0 \\
        0 & 0 & \lambda_3 & a_3 & 0 \\
        0 & 0 & 0 & \lambda_4 & a_4 \\
        0 & 0 & 0 & 0 & \lambda_5 
      \end{array}
    \right)
\]
where each $a_i$ is 0 or 1, and where, if $a_k = 1$ for some $k$,
then $\lambda_k = \lambda_{k+1}$.

For definiteness, suppose that $a_1 = 1$, $a_2 = a_3 = a_4 = 0$, and
hence that $\lambda_1 = \lambda_2 = \lambda$. Then there exist only
one eigenvector corresponding to this eigenvalue, let us call it
${\bf x}$: $M {\bf x} = \lambda {\bf x}$. But it is easy to see that
there exist another vector ${\bf y}$ (simply the second column in the
transformation matrix $P$) such that $M {\bf y} = \lambda {\bf y} +
{\bf x}$. Hence, operating with $M$ on any vector in the subspace
spanned by ${\bf x}$ and ${\bf y}$, yields a vector in the same
subspace. Therefore one says that ${\bf x}$ and ${\bf y}$ span a 2
dimensional invariant subspace corresponding to eigenvalue $\lambda$. Similarly,
if $a_1 = a_2 = 1$, $a_3 = a_4 = 0$ and hence $\lambda_1 = \lambda_2 =
\lambda_3$, $M$ would have a 3 dimensional invariant subspace.

From this we conclude that $\omega^a_{\;b}$ can be uniquely specified,
up to similarity, by its eigenvalues {\it and} the dimensions (and
signatures) of its invariant subspaces corresponding to these, because
this information is sufficient to fix the matrix $N$.

To actually perform this classification is tedious. However, the
observation that if $\lambda$ is an eigenvalue to $\omega^a_{\;b}$,
then $-\lambda$, $\lambda^*$ and $-\lambda^*$ are also eigenvalues,
greatly facilitates the calculation. For this, and for an explicit
demonstration of the algebraic manipulations, see the appendix of ref.
\cite{BH} where this classification is performed for SO(2,2). We quote
their result in table~1, and the corresponding result for SO(2,3) in
table~2. For the rest of this section we will compare and explain some
features of these tables.

The second and third columns give the characteristics of each type:
The eigenvalues (in terms of the real parameters $a$ and $b$) and the
dimension of the invariant subspaces, respectively. The latter is
directly reflected in the numbering of the types. The last column
provides an explicit Killing field, given values of the $a$:s and
$b$:s. Note that, because of these parameters, each type contains a
0-, 1- or 2-parameter family of non-equivalent fields.

Comparing the tables, we see that there are two more types in the
SO(2,3) case: Types $I_d$ and $V$. It is clear that they cannot exist
for SO(2,2) because they demand 3 spatial dimensions---all of the three
spatial indices 2, 3 and 4 occur in their Killing fields. Further, for
types $I\!I\!I_a$ and $I\!I\!I_b$ there is a 1-parameter family of
Killing fields in table~2. When the parameter is zero they both
coincide with the corresponding (0-parameter) fields in
table~1.\footnote{In ref. \cite{BH} types $I\!I\!I_a$ and $I\!I\!I_b$ are referred
  to as $I\!I\!I^+$ and $I\!I\!I^-$ respectively, because for SO(2,2)
  they only differ by the signature of the invariant subspace.}

As a curiosity, note that neither SO(2,2) nor SO(2,3) can have a 4
dimensional invariant subspace. When one performs the classification that turns
out to be inconsistent, in both cases, with the signature of the metric.

\begin{table}[p]
\[
\begin{array}{llll} \hline\hline
  {\it Type} & {\it Eigenvalues} & {\it Inv. subspace} & {\it Killing
  field}   \\ \hline \\[-3mm]
  I_a   & \!\!\!
  {\small
    \begin{array}{l}
      \lambda, -\lambda, \lambda^*, -\lambda^* \\
      \lambda = a + ib, a \neq 0, b \neq 0
    \end{array}}
  & \hspace{5mm} -\!\!\!-\!\!\!- & b(J_{01} + J_{23}) - a(J_{03} +
  J_{12}) \\[3mm]
  I_b & a_1, -a_1, a_2, -a_2 & \hspace{5mm} -\!\!\!-\!\!\!- &
  a_1 J_{12} + a_2 J_{03} \\[3mm] 
  I_c &ib_1, -ib_1, ib_2, -ib_2 & \hspace{5mm} -\!\!\!-\!\!\!- &
  b_1 J_{01} + b_2 J_{23} \\[3mm]
  I\!I_a & a, -a & {\small
    \begin{tabular}{l}
      2 dim. to $a$ \\
      2 dim. to $-a$
    \end{tabular}} & a (J_{03} + J_{12}) + J_{01} - J_{02} - J_{13} +
      J_{23} \\[3mm]
  I\!I_b & ib, -ib, & {\small
    \begin{tabular}{l}
      2 dim. to $ib$ \\
      2 dim. to $-ib$
    \end{tabular}} & (b-1) J_{01} + (b+1) J_{23} + J_{02} - J_{13} \\[3mm]
  I\!I\!I_a & 0 & {\small
    \begin{tabular}{l}
      3 dim. to $0$ \\
      with sign. $(++-)$
    \end{tabular}} & J_{23} - J_{03} \\[3mm]
  I\!I\!I_b & 0 & {\small
    \begin{tabular}{l}
      3 dim. to $0$ \\
      with sign. $(+--)$
    \end{tabular}} & J_{02} - J_{01} \\[3mm]
  \hline\hline
\end{array}
\]
\caption{Classification of SO(2,2)}  
\end{table}
\begin{table}[p]
\[
\begin{array}{llll} \hline\hline
  {\it Type} & {\it Eigenvalues} & {\it Inv. subspace} & {\it Killing
  field}   \\ \hline \\[-3mm]
  I_a   & \!\!\!
  {\small
    \begin{array}{l}
      \lambda, -\lambda, \lambda^*, -\lambda^*, 0 \\
      \lambda = a + ib, a \neq 0, b \neq 0
    \end{array}}
  & \hspace{5mm} -\!\!\!-\!\!\!- & b(J_{01} + J_{23}) - a(J_{03} +
  J_{12}) \\[3mm]
  I_b & a_1, -a_1, a_2, -a_2, 0 & \hspace{5mm} -\!\!\!-\!\!\!- &
  a_1 J_{12} + a_2 J_{03} \\[3mm] 
  I_c &ib_1, -ib_1, ib_2, -ib_2, 0 & \hspace{5mm} -\!\!\!-\!\!\!- &
  b_1 J_{01} + b_2 J_{23} \\[3mm]
  I_d & a, -a, ib, -ib, 0 & \hspace{5mm} -\!\!\!-\!\!\!- & a J_{03} +
  b J_{24} \\[3mm]
  I\!I_a & a, -a, 0 & {\small
    \begin{tabular}{l}
      2 dim. to $a$ \\
      2 dim. to $-a$
    \end{tabular}} & a (J_{03} + J_{12}) + J_{01} - J_{02} - J_{13} +
      J_{23} \\[3mm]
  I\!I_b & ib, -ib, 0 & {\small
    \begin{tabular}{l}
      2 dim. to $ib$ \\
      2 dim. to $-ib$
    \end{tabular}} & (b-1) J_{01} + (b+1) J_{23} + J_{02} - J_{13} \\[3mm]
  I\!I\!I_a & a, -a, 0 & {\small
    \begin{tabular}{l}
      3 dim. to $0$ \\
      with sign. $(++-)$
    \end{tabular}} & -a J_{14} + J_{23} - J_{03} \\[3mm]
  I\!I\!I_b & ib, -ib, 0 & {\small
    \begin{tabular}{l}
      3 dim. to $0$ \\
      with sign. $(+--)$
    \end{tabular}} & -b J_{34} + J_{02} - J_{01} \\[3mm]
  V & 0 & {\small
    \begin{tabular}{l}
      $\!\!\!\!$ { 5 dim. to} $0$
      \end{tabular}} & - J_{01} - J_{03} - J_{12} - J_{14} +
  J_{23} + J_{34} \\[1mm]
  \hline\hline 
\end{array}
\]
\caption{Classification of SO(2,3)}  
\end{table}

\section{Horizons in the quotient space}
In this section we will examine the causal structure for all quotient
spaces $[adS]/{\cal G}_{_\Gamma}$ systematically by choosing the
generator $\xi$ of $\Gamma$ from all different types found in the last
section. To do this we will use the tools developed in section~2, were
we defined the singularity and the horizons in terms of the
hypersurface $\xi^\mu\xi_\mu = 0$.

First, however, we will introduce a set of coordinates
that will enable us to visualize our conclusions. Since we wish to be
able to understand horizons and such, the best would be to find some
Penrose-like coordinates, that is, coordinates in which infinity
is mapped into a finite distance and where all lightlike curves have
slope 1. Unfortunately, in general there does not exist such
coordinates in higher dimensions than 1+1, and since we want
coordinates for 3 and even 4 spacetime dimensions we have to relax one
of these requirements.\footnote{We will choose coordinates which meet
  the requirement of mapping all points in spatial infinity into a
  finite distance. For a different choice, which fails to satisfy
  this, but where the other requirement is met instead---that the
  slope of null curves is 1---see ref. \cite{Brill}.}

In the spherical coordinates given below---$t$, $\rho$, $\phi$ and (in
4 dimensions) $\theta$---spacelike infinity is mapped into the
cylindrical surface $\rho = 1$. On the other hand the slope of the
null curves differ with the radius $\rho$, but for a given
point each null curve through that point has the same slope
irrespectively of its direction.
\begin{equation}
  \label{sausagecoord}
    \begin{array}{lll}
          X & = &  
          {\displaystyle \frac{2\rho}{1-\rho^2}}\sin{\theta}\cos{\phi}
          \\[4mm]
          Y & = &  
          {\displaystyle \frac{
          2\rho}{1-\rho^2}}\sin{\theta}\sin{\phi} \\[4mm]
          Z & = &  
          {\displaystyle \frac{2{\rho}}{1-{\rho}^2}}\cos{\theta}
          \\[4mm]
          U & = &  
          {\displaystyle \frac{1+{\rho}^2}{1-{\rho}^2}}\cos{t} \\[4mm]
          V & = &  
          {\displaystyle \frac{1+{\rho}^2}{1-{\rho}^2}}\sin{t}
    \end{array} 
\end{equation}
This parametrizes the defining hypersurface (\ref{hyperboloid}) for
3+1 adS-space. The radial coordinate $\rho$ takes values between 0 and
1 while $\phi$ and $\theta$ are the usual azimuthal and polar angles,
respectively. Since we are going to discuss causal behaviour we will
think of the time as ``unwound'', that is $-\infty < t < \infty$.
Otherwise we would have closed timelike curves already from the start.
(Formally this means that we are actually working with the universal
covering of adS-space.) The metric becomes
\begin{equation}
  \label{metric}
    ds^2 = - \left(\frac{1 + {\rho}^2}{1 - {\rho}^2}\right)^2 dt^2 +
           \frac{4}{(1 - {\rho}^2)^2}(d{\rho}^2 + {\rho}^2d{\theta}^2
           + {\rho}^2\sin^2{\theta}d{\phi}^2) \ 
\end{equation}

First, let us concentrate on the 2+1 dimensional case for which we
simply put $\theta = \frac{\pi}{2}$ in equations (\ref{sausagecoord})
and (\ref{metric}).  Then these coordinates describe adS-space as the
interior of an infinitely long cylinder whose surface $\rho = 1$
represents spatial infinity. Every constant time slice of this
cylinder yields the Poincar\'{e} disk model of a
negatively curved surface of infinite area, which among other things means that the
geodesics in such slices are arcs of circles meeting the boundary
$\rho = 1$ at right angles.

Because of our ``unwinding'' of the time parameter all timelike
geodesics exhibit a periodicity of $2\pi$, generically spiraling
through the cylinder. Each lightlike geodesic start and end in
spatial infinity during a finite (coordinate) time interval. In
particular, it takes a time $\pi$ for a lightray to travel from one
side of the cylinder, through the origin $\rho = 0$, to the opposite
side. This behaviour is expected since, as is well known, adS-space
fails to be globally hyperbolic; information keeps flowing in from
infinity and there do not exist any Cauchy surfaces. Actually, it is
this property of adS-space that renders possible the appearance of
interesting causal structures in its quotient spaces.

In order to understand the 3+1 dimensional case we just have to think
about each constant time slice of the cylinder as, not a Poincar\'{e}
disk, but a Poincar\'{e} ball. The geodesics in this ball still
consist of all arcs of circles meeting its boundary at right angles.

Now, at last, we are ready to investigate the causal structure of
$[adS]/{\cal G}_{_\Gamma}$ where the generator to $\Gamma$ belongs to
one of the possible Killing field types found in the last section. In
order to make the procedure clear we will describe the first case in
some detail, but then merely state the results together with some
comments.

\vspace{3mm}
\noindent{\underline{\bf Type $\bf I_a$}}

\noindent Using equation (\ref{baseKilling}) to write the Killing
field from table~1 or 2 in terms of the embedding coordinates, we have
\[
\xi^\mu = ( -aY - bV) \frac{\partial}{\partial U} + (-aX + bU)
\frac{\partial}{\partial V} + (-aV + bY) \frac{\partial}{\partial X} + (-aU - bX)
\frac{\partial}{\partial Y}
\]
for both 4 and 5 dimensions. Using equation (\ref{hyperboloid}) this
gives the norm
\[
\xi^\mu \xi_\mu = (a^2 - b^2) (Z^2 + 1) - 4ab (YV - UX)
\]
\vspace{3mm}
\begin{figure}[h]
\hfil
\begin{minipage}[b]{75mm} 
\caption{The type $I_a$ singularity surface (for a = b) is shown
during a time interval $\Delta t = \pi$ using the cylinder coordinates
$\rho$, $\phi$ and $t$ introduced in the text. The
boundary of the cylinder represents spatial infinity. The type $I_a$
Killing field is timelike on one side of this corcscrew like surface,
and spacelike on the other.}
\end{minipage}
\hfil
\begin{minipage}[b]{60mm}
  \epsfbox{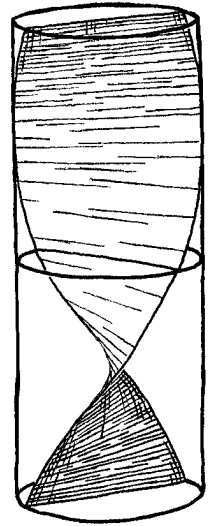}
\end{minipage}
\hfil
\end{figure}
from which we see that the Killing field is everywhere timelike when
$a = 0, b \neq 0$, and everywhere spacelike when $a \neq 0, b = 0$.
However, when both $a$ and $b$ are non-zero we get both a timelike and
a spacelike region and thus a singularity surface $\xi^\mu \xi_\mu =
0$ in between, according to our terminology from section~2.

In figure~1 we have drawn this singularity in the special case $a = b$
for 2+1 adS-space using the coordinates introduced above. It is a
corkscrew shaped surface meeting the infinity in two lines spiraling
around the cylinder; these are lightlike geodesics as they have to be
according to the analysis in section~2. When $a \neq b$ these lines
remain the same, but the singularity between them then bulge in one
or the other direction, making either the timelike or the spacelike
region larger.

Since, in the part where $\xi$ is spacelike, infinity is connected
and remains for all $t$:s, it is always possible for an observer to
escape to it. Thus there does not exist an event horizon in this case.
Since the singularity is timelike, it is visible everywhere and
therefore there is no inner horizon either.

These results extend to the 3+1 dimensional case. Then the singularity
in each Poincar\'{e} ball (that is, at each time $t$) consists of a
surface again dividing it into one timelike and one spacelike part.
This singularity only rotates in time around the axis $\theta = 0$,
and hence, for the same reasons as before, there can be no horizons in
this case either.

\vspace{10mm}

\begin{figure}[h]
\hfil
\begin{minipage}[b]{120mm} 
  \epsfbox{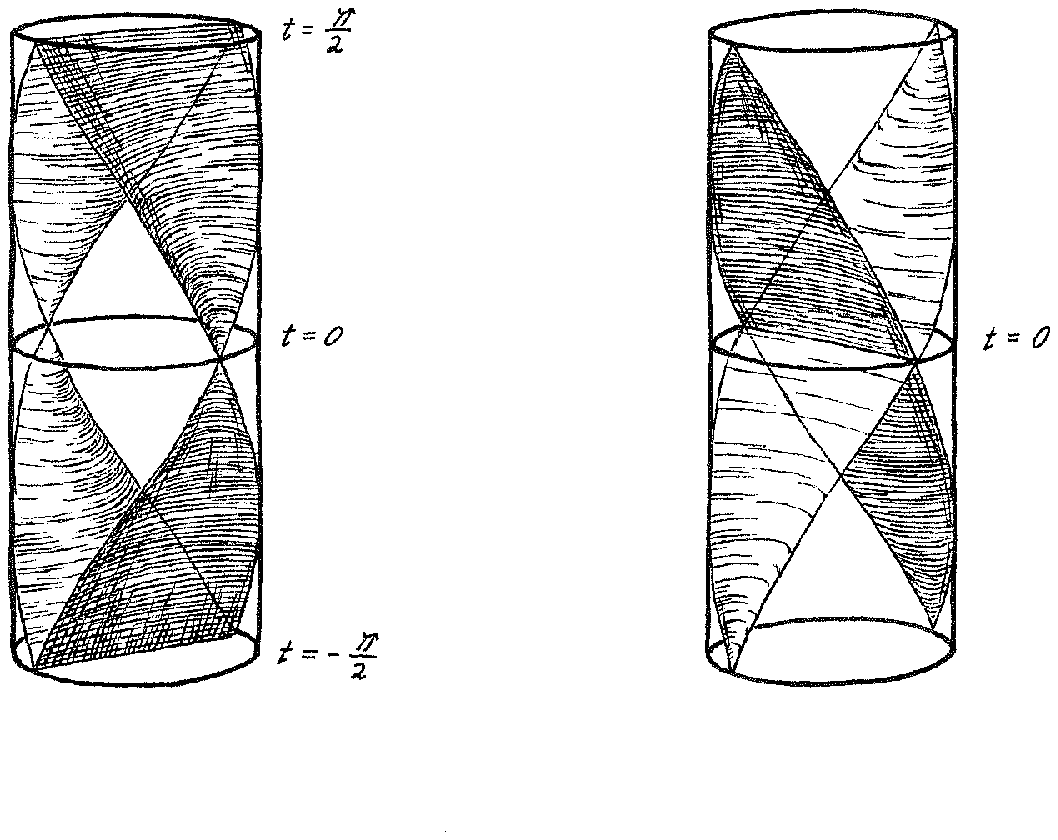}
  \vspace{-5mm}
  \caption{The static BTZ black hole: Type $I_b$ with
$a_1 = 0$ (or $a_2 = 0$).  Regions of spacelike Killing field last
only for finite coordinate time intevals $\pi$. Such a region is
bounded by spatial infinity and the singularity surface shown in the
left adS-cylinder. The corresponding event horizon is depicted to the
right.}
\end{minipage} 
\hfil
\end{figure}

\vspace{3mm}
\noindent{\underline{\bf Type $\bf I_b$}}

\noindent Let us consider the 2+1 dimensional case first. This choice
for $\xi$ then yields the BTZ black hole. When either $a_1$ or $a_2$
equals zero we get the static non-rotating case, that is, only then
the quotient space admits a timelike hypersurface orthogonal Killing
field. The singularity surface---described by $U = Y$ if we put $a_1 =
0$---is depicted to the left in figure~2. We see that at the time symmetric
moment $t = 0$ it consists of two opposite points on the border of the
cylinder, that is, on the infinity. From these points it grows up forward
and backward in time, at each instant consisting of two opposite
geodesics. These grow larger and larger until they come together at $t
= \pm \frac{\pi}{2}$. Therefore the relevant region
$\xi^\mu \xi_\mu > 0$ only exists for a finite coordinate time
interval $\Delta t = \pi$, and not all of it is visible from
infinity. Thus, there exists an event horizon which turns out to be
described by $V = X$. It is the same type of surface as the
singularity, only rotated $\pi/2$ around the symmetry axis of the
cylinder and translated $\pi/2$ in coordinate time, see the right
cylinder in figure~2. However, it is not possible for an observer to see the singularity
before he hits it, and hence there is no inner horizon.

When both $a_1$ and $a_2$ are non-zero one obtains the more general
rotating BTZ black hole. As is seen from figure 3 the singularity
\begin{figure}[h]
\hfil
\begin{minipage}[b]{70mm} 
\caption{When both $a_1$ and $a_2$ are non-zero type $I_b$ yields the rotating
BTZ black hole. The singularity surface opens up at $t = \pm
\frac{\pi}{2}$, and the spacelike Killing field region is not
interrupted any longer at these coordinate times. As a result, there
are not only event horizons but also inner horizons.}
\end{minipage}
\hfil
\begin{minipage}[b]{60mm}
  \epsfbox{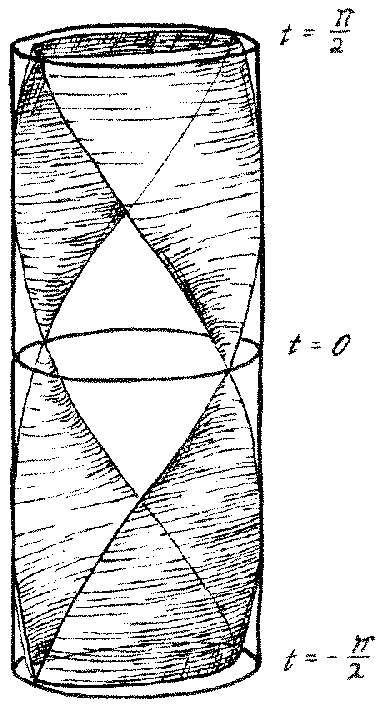}
\end{minipage}
\hfil
\vspace{5mm}
\hfil
\begin{minipage}[c]{140mm} 
  \epsfbox{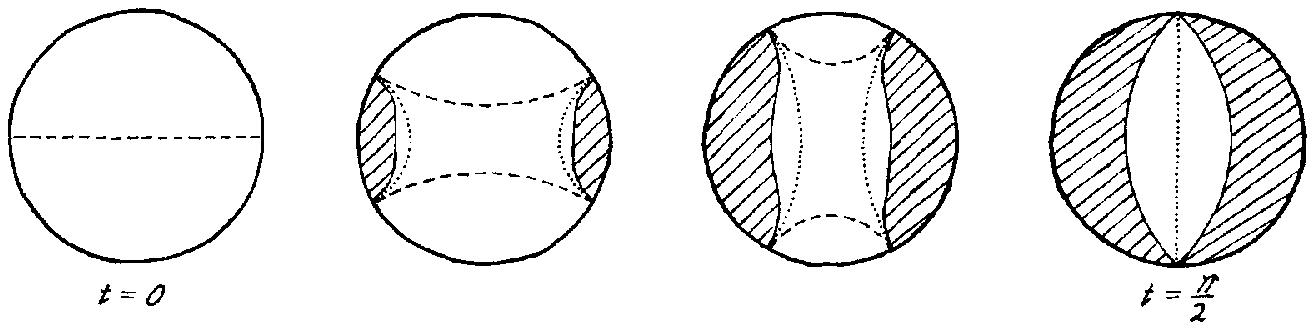}
  \vspace{-6mm}
  \caption{The rotating BTZ black hole as a time series of Poincar\'{e} disks.
The shaded regions indicates where the Killing field is timelike and
may be thought of as excluded from the spacetime. The event horizon
is dashed, while the inner horizon is dotted.}
\end{minipage} 
\hfil
\end{figure}
surfaces do not meet each other at $t = \pm \pi/2$ any longer; an
observer sitting in the center of the cylinder never hits the
singularity. Therefore the quotient space will contain not only event
horizons, one for each connected part of infinity, but also inner
horizons, one for each singularity surface. Since each connected part
of infinity is the same as in the non-rotating case, each event
horizon again will be the surface depicted to the right in figure~2. The inner horizons
happens to look the same as the singularity surface in the non-rotating
case, to the left in figure~2. For clarity, both the singularity and the two
horizons are drawn in figure~4 in a time series beginning with the
time symmetrical Poincar\'{e} disk at $t = 0$.

Let us now see how these results generalize to 3+1 dimensions. The
singularity surface in the non-rotating case, together with the
resulting event horizon, is depicted in the time series in figure~5.
This series could be thought of as obtained simply by rotating the
Poincar\'{e} disk at each instant for the 2+1 non-rotating case
(figure 2) in such a way that the opposite singularity geodesics, or
arcs of circles, instead becomes segments of spheres; the horizon becomes
the cylindrical membrane in the figure, connecting these spheres.
Note, however, that in the quotient space this cylindrical horizon
actually is a torus, on account of the identification.  Superficially
all this seems very similar to the 2+1 case: The singularity surfaces
again meet at $t = \pm \pi/2$, and the event horizon grows up from $t
= 0$. But a closer analysis reveals some important differences. Most
notably this black hole turns out to be non-static and to possess a
non-trivial apparent horizon; these properties will be discussed
further in the next section. 

What is the 3+1 counterpart to the rotating BTZ black hole? 
As in the 2+1 case the singularity starts out at $t = 0$ from two
oppositely lying points at infinity. The two surfaces growing up
from these points move inwards as time passes until they reach their
closest position at $t = \pi/2$, after which they recede from each
other again. At this moment, depicted in figure~6, they touch
in two points, lying opposite to each other at infinity. In 2+1
dimensions this ``touching'' is enough to make the infinity
disconnected, giving rise to one event horizon for each part of it.
Here, on account of the third space dimension, the infinity remains
connected, and therefore there will be no event horizons. Thus, {\it
  it is not possible to generalize the rotating BTZ-black hole to 3+1
  dimensions, because the analogue does not contain any event
  horizons.}

However, there is an inner horizon for each part of the singularity,
which looks the same as the singularity surface in figure~5.  

\vspace{5mm}
\begin{figure}[h]
\hfil
\begin{minipage}[c]{140mm} 
  \epsfbox{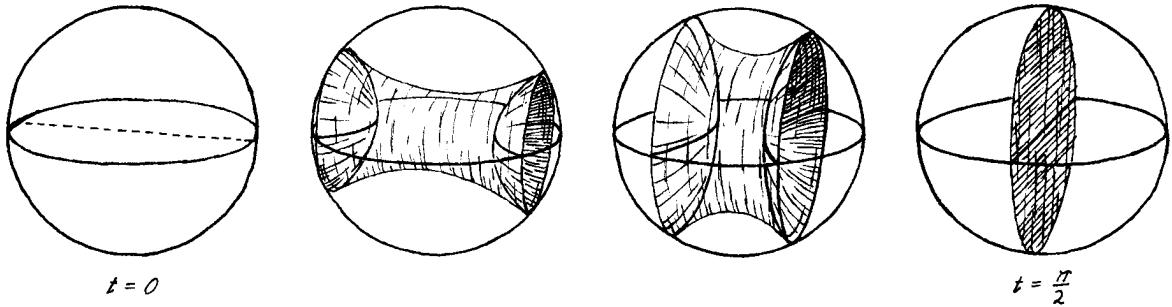}
  \vspace{-3mm}
  \caption{The 3+1 dimensional version of the non-rotating BTZ hole shown as a
time series of Poincar\'{e} balls. The opposite spherical segments
are the singularity, and the cylindrical membrane connecting these is
the event horizon.}
\end{minipage} 
\hfil
\end{figure}
\begin{figure}[h]
\hfil
\begin{minipage}[b]{60mm} 
\caption{The Poincar\'{e} ball at $t = \pm \frac{\pi}{2}$ for the 3+1 version
of the rotating BTZ hole. At this moment the singularity surfaces have
reached their closest position, but there is still infinity left in
between. This is in contrast to the situation in the 2+1 dimensional
counterpart, the last Poincar\'{e} disk in the series of figure~4.}
\end{minipage}
\hfil
\begin{minipage}[b]{60mm}
  \epsfbox{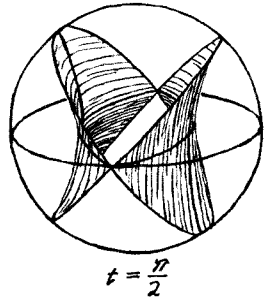}
\end{minipage}
\hfil
\end{figure}

\vspace{3mm}
\noindent{\underline{\bf Type $\bf I_c$}}

\noindent The singularity surface only exists when both $b_1$ and
$b_2$ are non-zero and $|b_2| > |b_1|$. It turns out to be independent
of coordinate time $t$, and therefore looks the same in every constant
time Poincar\'{e} disk or ball. In 2+1 dimensions it simply is a
circle centered around $\rho= 0$ in the Poincar\'{e} disk whose radius
depends on $b_1$ and $b_2$. In 3+1 dimensions it is a cylinder
centered around the axis $\theta = 0$ (or $\pi$) in the Poincar\'{e}
ball. There do not exist any horizons in either case.

\vspace{3mm}
\noindent{\underline{\bf Type $\bf I_d$}}

\noindent This case only exists in 3+1 dimensions. For $b = 0$ it
coincides with the non-rotating BTZ black hole (type $I_b$ with $a_1 =
0$). As in that case the singularity surfaces grows up from two
oppositely lying points in the infinity at $t = 0$. However, for
non-zero $b$, these surfaces do not come together in $t = \pi/2$; they
merely touch at $\rho = 0$. In particular, the infinity is connected
and exist for all times $t$, and hence there will be no event
horizons. But there will be inner horizons, again of the same shape as
the singularity in figure~5, one for each of the disconnected
singularity surfaces.

\vspace{3mm}
\noindent{\underline{\bf Type $\bf I\!I_a$}}

\noindent When $a = 0$ this Killing field is null everywhere, and we
can not talk about a singularity surface at all. For $a \neq 0$ it
gives rise to what Ba\~nados {\it et al} \cite{Banados,BH} call ``the
extremal $J = M$ black hole'', because in their setting it appears as
a natural limit of the general type $I_b$ black hole. However,
actually there are no event horizons in this case since the infinity
is connected and lasts forever in coordinate time. On the other hand,
the very complicated looking singularity surface consists of
disconnected parts, and so there will be inner horizons.

This terminology of calling something a black hole that actually is
not may seem strange, but we remind the reader that this is
the situation also for the asymptotically anti-de Sitter extremal Kerr
black hole. This is not a black hole either, in the sense of the
definition, due to the null infinity of adS-space being a ``vertical
line'' (referring to its appearence in a Penrose diagram).
Namely, in the extremal Kerr solution there is
nothing that interupts this vertical line infinity, and so there can
be no event horizons. Likewise, one could argue that the usual
extremal Kerr black hole, that is, the asymptotically flat one, is a
real black hole only because of the structure of Minkowskian infinity,
consisting of a past and a future null infinity. This makes the
infinity for the extremal Kerr solution being a ``zig-zag line''
consisting of alternate future and past infinities, hence giving rise
to event horizons.

Note, therefore, that this type provides an example of a case where
the infinity structure of adS-space makes it {\it harder} to form a
black hole. This is contrary to the impression given by the fact that
the BTZ-hole exists only in adS-background---its counterpart in
Minkowski background, the Misner space, contains no horizons.

\vspace{3mm}
\noindent{\underline{\bf Type $\bf I\!I_b$}}

\noindent There are no horizons in this case either. In 2+1 dimensions, at each moment,
the singularity is a loop in the Poincar\'{e} disk touching infinity
at one point. This loop whirls around as time passes without changing
its shape. When the third spatial dimension is included the loop
becomes a rather complicated surface in the Poincar\'{e} ball.
However, its time dependence still only consists of a ``rigid
rotation'' around the axis $\theta = 0$ (or $\pi$) and thus, since the
singularity has to be a timelike surface by the general arguments in
section~2, it can not give rise to any horizons.

\vspace{3mm}
\noindent{\underline{\bf Type $\bf I\!I\!I_a$}}

\noindent For $a = 0$ this type exists for both 2+1 and 3+1
dimensions.  In the former case it gives rise to the extremal $M = 0$
black hole in the terminology introduced by Banados {\it et al}
\cite{BH}; the singularity is the null surface shown in figure~7.
\begin{figure}[h]
\hfil
\begin{minipage}[b]{60mm}
\caption{The 2+1 dimensional type $I\!I\!I_a$ singularity. The Killing field
actually is spacelike on both sides of this surface, but it has fix
points (that is, it equals zero) along the dashed lines in the figure.}
\vspace{7mm}
\end{minipage}
\hfil
\begin{minipage}[b]{60mm}
  \epsfbox{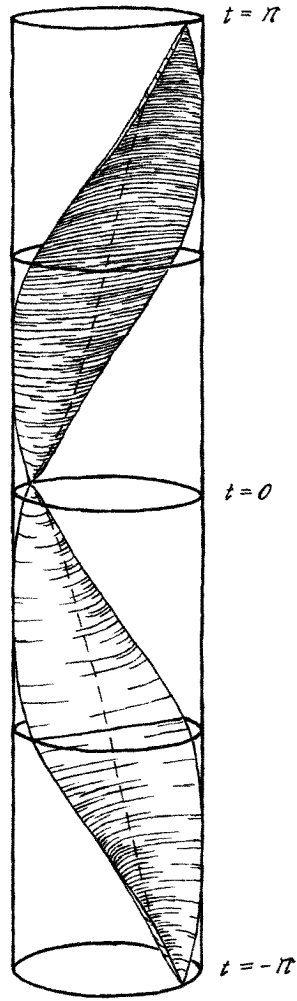}
\end{minipage}
\hfil
\end{figure}
Actually, $\xi^\mu \xi_\mu > 0$ on both sides of this surface, and
therefore there is no region in the quotient space containing closed
timelike curves. However, it still makes sense to call it a
singularity, because it consists of closed null lines, and furthermore,
$\xi$ has a line of fixpoints at $Y = 0$, $X = -U$ (the dashed line in
the figure) making the quotient space singular there (non-Hausdorff,
to be precise).

It is always possible for an observer to escape to infinity, so there
is no event horizon. It is not possible for an observer to see the
future part of the singularity before he hits it, while the past part
is visible from everywhere. Thus there is no inner horizon either.

The extremal $M = 0$ BTZ black hole easily generalizes to 3+1
dimensions. Just take every constant time Poincar\'{e} disk in
figure~7 and rotate it to obtain the Poincar\'{e} ball, in such a way
that the singularity becomes a segment of a sphere instead of an arc
of a circle. This sphere segment then bounces to-and-fro in the
Poincar\'{e} ball in time. For the same reasons as before there are no
horizons.

The case $a \neq 0$ only exists in the 3+1 case. The singularity then
consists of two ellipsoidal segments bouncing around in the
Poincar\'{e} ball. Again, there is no event horizon. But since these
segments, with a period of $\Delta t = \pi$, shrinks into two points
at infinity, the singularity consists of several disconnected
parts. To each of them there is a corresponding inner horizon, again
with the shape of the singulatity surface in figure~5.

\vspace{3mm}
\noindent{\underline{\bf Type $\bf I\!I\!I_b$}}

\noindent In the 2+1 case, that is, when $b = 0$, the Killing field is
everywhere timelike and there is closed timelike lines through every
point in the quotient space. When $b \neq 0$ we get the singularity
surface illustrated in figure~8, which, despite its fancy appearance,
does not give rise to any horizons.

\vspace{3mm}
\noindent{\underline{\bf Type $\bf V$}}

\noindent This 3+1 dimensional type does not give rise to any horizons
either. The singularity could be described as a sheet with a kink
rocking up and down in the Poincar\'{e} ball.

\begin{figure}[h]
\hfil
\begin{minipage}[c]{140mm} 
  \epsfbox{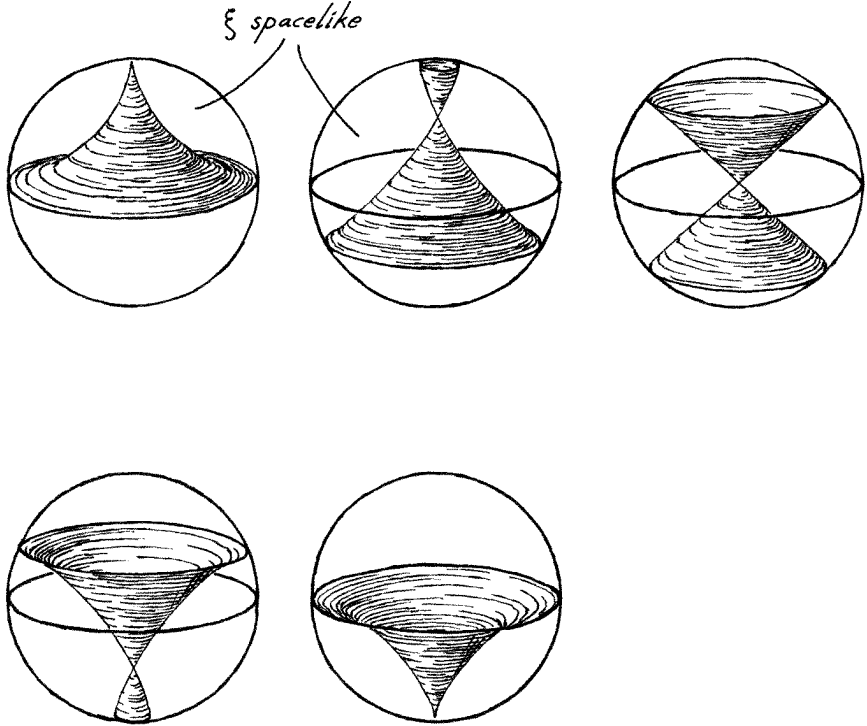}
  \vspace{-3mm}
  \caption{The type $I\!I\!I_b$ ($b \neq 0$) singularity. This series show the
singularity during an interval $\Delta t = \pi$. Having reached the
position in the last picture the surface oscillates back again
following this time series backwards.}
\end{minipage} 
\hfil
\end{figure}

\section{A closer look at the 3+1 black hole}
In the previous section we found and catalogized all possible causal
structures in spacetimes of the form $[adS]/{\cal G}_{_\Gamma}$. In the 3+1
dimensional case, although we did find some strange looking
singularities which in many cases gave rise to inner horizons, there
was only one case containing an event horizon: Type $I_b$ when either
$a_1$ or $a_2$ vanishes. Actually, the most surprising conclusion we
could draw was a negative one: There does {\it not} exist any 3+1
version of the rotating BTZ black hole, because the third space
dimension provides extra infinity to which an observer always might
escape. Thus, it is only the non-rotating case that admits a 3+1
analogue. However, this analogue has some interesting properties which
it does not share with the 2+1 case. In this section we will
investigate these new features.

First, we may ask what are the symmetries of our black hole. When
making the identification $\Gamma = e^{\alpha \xi}$ in adS-space we lose all
Killing fields that are not single valued in the quotient space, that
is, those fields that are not mapped into themselves under $\Gamma$.
This means that the remaining Killing fields will be exactly those
which commute with $\xi$ \cite{BH}. Therefore $\xi$ itself obviously has to
be a symmetry of the quotient space, and, since both SO(2,2) and
SO(2,3) have rank 2, at least one more Killing field should survive
the identification.

For our non-rotating black holes $\xi$ equals $J_{03}$, putting $a_1 =
0$ for definiteness. Considering the 2+1 case first, the only field
commuting with $J_{03}$ apart from itself is $J_{12}$, as is easily
shown from equation (\ref{baseKilling}). This field is timelike for $X
> V$, which is---recalling from section~4 that $X = V$ is the event
horizon---precisely the exterior region. Furthermore, $J_{12}$ is
hypersurface orthogonal, and thus the 2+1 case must be static,
describing an eternal black hole. In particular, the area of the event
horizon does not change with time, and can be shown to be
$(\frac{\alpha}{2\pi})^2$ (for $\Gamma = e^{\alpha J_{03}}$).

The situation for the 3+1 black hole is different. It follows from
equation (\ref{baseKilling}) that the Killing fields commuting with
$J_{03}$ apart from itself are $J_{12}$, $J_{14}$ and $J_{24}$, so
these then are our symmetries. Of these $J_{12}$, $J_{14}$ and
$J_{24}$ are generators of the 2+1 dimensional Lorentz group SO(1,2),
something that is most easily seen from their explicit form
(\ref{baseKilling}).  Hence, since the fourth symmetry $J_{03}$
commutes with all of them, the full symmetry group for our 3+1 black
hole must be ${\rm SO(1,2)} \times {\rm U(1)}$.

The event horizon---depicted in figure~5 as the cylindrical membrane
connecting the growing singularity surface segments---is given by $V^2
= X^2 + Z^2$. In contrast with the 2+1 case, there is no Killing field
that is timelike everywhere in the exterior region 
\begin{equation}
  \label{ExteriorRegion}
  V^2 < X^2 + Z^2
\end{equation}
Indeed, the only candidates would be of the form $J_{12} \cos \chi +
J_{14} \sin \chi$, since $J_{03}$, the identification field itself, is
by construction spacelike in the interesting region, whereas
$J_{24}$ is a pure rotation, hence also spacelike. But this
combination is timelike only where
\begin{equation}
  \label{timelikeKilling}
  |V| < |X \cos \chi + Z \sin \chi|  
\end{equation}
For a given $\chi$, this inequality does not cover the whole exterior
region; only for $Z = X \tan \chi$ it is fulfilled all the way in to
the event horizon. Hence, although the union of all regions
(\ref{timelikeKilling}) for different values of $\chi$ covers precisely
the exterior region, there do not exist any Killing field that by
itself is timelike everywhere outside the event horizon, and we may
conclude that the black hole is non-static. Equivalently, we could say
that the horizon is not a Killing horizon, that is, its generators do
not belong to one single Killing field.

Intuitively this may be seen already from figure~5. The circumference
of the cylindrical horizon obviously increases in time, while its
length must be the same as the length of the event horizon in the 2+1
case: $(\frac{\alpha}{2\pi})^2$ and constant in time. Thus
its area is increasing in time, and the spacetime has to be
non-static. Moreover, the fact that the horizon is growing only in one
of its directions shows that it must be shearing.
 
As noted already in the previous section the horizon actually is a 
torus on account of the identifiaction, and the spacetime could be 
said to describe a growing toroidal black hole. The reader may feel
suspicious against this non-trivial topology of the horizon. But
actually, there is nothing else to expect, since the whole spacetime
has topology ${\bf T^2 \times R^2}$ and infinity itself is, at
each instant, a torus.

The observation that the event horizon is growing lead us to one
further question: Where is the apparent horizon? In the 2+1
dimensional BTZ-black hole this horizon coincides with the event
horizon (as discussed in ref. \cite{Carlip} and \cite{ViOchBrill}).
This actually has to be the case for all static black holes. However,
by a theorem of Hawking and Ellis \cite{HawkingEllis} (carefully
discussed and proved in ref.  \cite{Kriele})---which states that the
apparent horizon must be marginally trapped---it cannot coincide with
a growing event horizon.  Thus, our 3+1 black hole, despite its
trivial nature of being locally anti-de Sitter, must contain a
non-trivial apparent horizon. After defining some relevant concepts we
will devote the rest of this section to finding its location.

First, a {\it trapped surface} is a spacelike surface such that, for
each of its points, the two families of lightrays orthogonal to it
converge in their future direction. Actually, there is nothing unusual
with such a surface; a particularly clear example is provided by the
intersection of two backward lightcones in Minkowski space: The two
families of lightrays orthogonal to this intersection are nothing else
than the lightcone generators themselves, which, of course, converge
to the top of the cones.

The interesting situation occurs when the trapped surface is closed,
that is, compact and without edge; such a surface is called a {\it
  closed trapped surface}. By the singularity theorems formulated and
proved in the sixties by Hawking and Penrose (se {\it e.g.} ref.
\cite{Wald}) such surfaces signal the existence of a future
singularity. Actually, in the proof of these theorems one assumes the
existence, not of closed trapped surfaces, but of the slightly
different concept of {\it closed outer trapped surfaces}. For these
only the outgoing family of lightrays converge.  Also, they
have to be the boundary to a volume, something that actually is {\it
  not} required for a closed trapped surface.  Finally, a {\it closed
  marginally outer trapped surface} is a closed outer trapped surface
except that the outgoing family of lightrays has zero convergence.

Now, consider a foliation $\Sigma$ of the spacetime. For each slice
$\Sigma_i$ in this foliation, the {\it apparent horizon} is defined as
the boundary to the region in $\Sigma_i$ containing closed outer
trapped surfaces. As already mentioned, the apparent horizon can be
shown to be marginally outer trapped. Another well-known result is
that it has to reside inside the black hole, that is, behind the event
horizon.

Note that this definition, which is the conventional one, is strongly
foliation dependent; the apparent horizons with respect to two
different foliations do in general not give the same 3-surface in
spacetime. One could of course define an ``invariant'' apparent
horizon as the boundary to the spacetime region containing {\it all}
closed outer trapped surfaces that exist in the spacetime. However, we
do not know of any theorems concerning such a horizon, and it would
probably be a very difficult task to actually locate it in a given
situation. Thus we will stick to the conventional definition.

The first thing to do then is to fix a suitable foliation. Our choice
will be governed by the requirement that it should preserve all
symmetries of the spacetime. This will actually make our choice
unique, and in order to find it, note that three of the
symmetries---$J_{12}$, $J_{14}$ and $J_{24}$---act as SO(1,2) on the
\{V,X,Z\}-subspace of the embedding space (\ref{5-metric}), hence
leaving the hyperboloids $X^2 + Z^2 - V^2 = c$ invariant. The fourth
symmetry $J_{03}$ only affects the \{U,Y\}-plane, and therefore these
hyperboloids are preserved by the whole symmetry group. For $c < 0$
they are spacelike and---recalling inequality (\ref{ExteriorRegion})
for the exterior region---cover precisely the interior of the  black
hole. This is enough, since we only expect to find closed trapped
surfaces inside the black hole anyway.

In order to make the foliation explicit, let us parametrize the
interior region with the so called Schwarzschild coordinates $r$, $t$,
$\varphi$, $\psi$ (due to the Schwarzschild like properties of the
resulting metric (\ref{SchwarzschildMetric})):
\begin{equation}
  \label{SchwarzschildCoord}
    \begin{array}{lll}
          X & = & -\sqrt{1-r^2} \sinh t \cos \varphi \\
          Y & = & r \sinh \psi \\
          Z & = & -\sqrt{1-r^2} \sinh t \sin \varphi \\
          U & = & r \cosh \psi \\
          V & = & -\sqrt{1-r^2} \cosh t
    \end{array} 
\end{equation}
Here $r \in (0,1)$ acts as a timelike coordinate, whereas $t \geq 0$
acts as a radial one (!). Coordinates $\varphi$ and $\psi$ are
angular variables, corresponding to the closed symmetries $J_{24}$ and
$J_{03}$, respectively. As a result of the identification with $\Gamma =
e^{\alpha J_{03}}$, $\psi$ takes values only between 0 and $2\pi
\alpha$, these being identified. (These coordinates may be extended to the
exterior region ($r > 1$) by changing sign inside and outside the
square root in $X$, $Z$ and $V$, and making the exchange $\sinh t
\leftrightarrow \cosh t$, see {\it e.g.} ref. \cite{BH}.) The metric
becomes
\begin{equation}
  \label{SchwarzschildMetric}
  ds^2 = - \frac{1}{1-r^2} dr^2 + (1-r^2) dt^2 + r^2 d\psi^2 + (1-r^2)
  \sinh^2 \!t\; d\varphi^2
\end{equation}
As the ordinary Schwarzschild metric this one is singular at $r = 0$
and $r = 1$. These values correspond to the singularity and the
horizon, respectively. The symmetry preserving foliation found above
is given simply by slices of constant $r$.

Now then, how do we find the apparent horizon with respect to this
constant $r$ foliation? The first thing to note is that, since our
slicing respects the spacetime symmetries---a property that makes it
unique---the apparent horizon clearly has to do that too. But this
means that it actually has to {\it coincide} with one of these
constant $r$ slices. The question is only for which value of $r$ this
happens.

\begin{figure}[h]
\hfil
\begin{minipage}[b]{60mm} 
\caption{A qualitative picture of the interior region clarifying the relation
between the Schwarzschild coordinates $r$ and $t$, and the lightlike
ones $u$ and $v$ used in the text to find the apparent horizon.}
\end{minipage}
\hfil
\begin{minipage}[b]{60mm}
  \vspace{10mm}
  \epsfbox{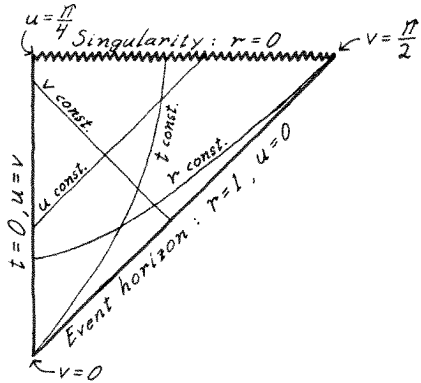}
\end{minipage}
\hfil
\end{figure}

In order to solve this problem we will use the method described for
example in ref. \cite{PetersBok}. First, introduce the lightlike
coordinates $u$ and $v$ expressed in the Schwarzschild parameters $r$
and $t$ as
\begin{equation}
  \label{uvCoordinates}
  \begin{array}{lll}
    u & = & {\displaystyle \arctan \left( e^{-t}
      \sqrt{\frac{1-r}{1+r}} \; \right)} \\[5mm]
    v & = & {\displaystyle \arctan \left( e^{t}
      \sqrt{\frac{1-r}{1+r}} \; \right)}
  \end{array}
\end{equation}
Since $0 < r < 1$ and $t \geq 0$ we have that $0 < u < \frac{\pi}{4}$
and $0 < v < \frac{\pi}{2}$. Directly from equations
(\ref{uvCoordinates}) we also have that $v > u$. In figure~9 the
meaning of $u$ and $v$, and their relation to $r$ and $t$, is
clarified. The metric (\ref{SchwarzschildMetric}) goes into
\begin{equation}
  \label{uvMetric}
  ds^2 = \frac{1}{\cos^2 (u-v)} ( -4dudv + \cos^2 (u+v) d\psi^2 +
  \sin^2 (u-v) d\varphi^2 )
\end{equation}

Next, consider the lightlike surfaces of constant $u$. Their lightlike
normal field is given by
\begin{equation}
  \label{uNormalField}
  l^\alpha = - \nabla^\alpha u
\end{equation}
The minus sign makes the field future directed. This means that
$l^\alpha$ could be viewed as the outward future directed lightlike
normals of 2-surfaces of constant $r$ and $t$. These surfaces, being
closed in $\psi$ and $\varphi$, are all toroidal, and, because of
their constant radius $t$, symmetric with respect to these variables.
We will find all such symmetric trapped surfaces first, and then argue
that taking into account non-symmetrical ones as well would not affect
the resulting apparent horizon.

The divergence $\vartheta$ of the outward directed lightrays
orthogonal to these constant \{$r$, $t$\} surfaces are then given by
\begin{equation}
  \label{Expansion}
  \vartheta = \nabla_\alpha l^\alpha = -\nabla_\alpha \nabla^\alpha u
\end{equation}
But, to be honest, this statement does not follow only from the
observation made above that $l^\alpha$ is the normal field to such
surfaces. Namely, the divergence of lightrays starting out from a
2-surface should only depend on its extrinsic curvature, or in our
case, how the normal field changes in the $\psi$ and $\varphi$
directions along the surface. But why does not the divergence
(\ref{Expansion}) acquire contributions from the $u$ and $v$
directions as well? To understand this, note that the metric
$\tilde{g}^{\alpha\beta}$ for an \{$u$,$v$\}-plane may be written
\[
\tilde{g}^{\alpha\beta} = f(u,v) (l^\alpha m^\beta + m^\alpha l^\beta)
\]
with $l^\alpha$ as in (\ref{uNormalField}) and $m^\alpha =
-\nabla^\alpha v$. Let ${\cal D}_\alpha$ denote the covariant
derivative with respect to this metric. Then
\[
{\cal D}_\alpha l^\alpha = \tilde{g}^{\alpha\beta} {\cal D}_\alpha
l_\beta = f(u,v) (l^\alpha m^\beta + m^\alpha l^\beta) {\cal D}_\alpha
l_\beta
\]
But since $l^\alpha$ is a gradient field ${\cal D}_\alpha l_\beta =
{\cal D}_\beta l_\alpha$, and we get
\[
{\cal D}_\alpha l^\alpha = 2 f(u,v) l^\alpha m^\beta {\cal D}_\beta
l_\alpha = 0
\]
where the last equality follows since $l^\alpha$ is null. Thus
the divergence of $l^\alpha$ gets no contribution from the $u$ and $v$
directions.

Having clarified the meaning of equation (\ref{Expansion}) we now
express it in terms of the Schwarzschild parameters $r$ and $t$ (a
calculation which is facilitated by changing $u$ and $v$ for $\tau = v
+ u$ and $\zeta = v - u$):
\begin{equation}
  \label{ExpansionEquation}
  (1-r^2)(\sinh 2t - 2r\cosh 2t) = 2r^3 - \vartheta
  r\sqrt{1-r^2}\sinh t (1 + (1-r^2) \sinh^2\!t\;)
\end{equation}
The divergence $\vartheta$ of a given closed 2-surface of constant
radius $t$ in the slice $r$ has to satisfy this relation. For this
surface to be trapped $\vartheta$ must be less than zero; $\vartheta =
0$ gives a marginally trapped surface. Note that $\vartheta \leq 0$ clearly
implies that the right-hand side of equation (\ref{ExpansionEquation}) is
positive. However, the left-hand side is negative for $\frac{1}{2} < r
< 1$, since $\cosh x > \sinh x$ for all $x$.Thus there can be no
trapped surfaces of this type for $r > \frac{1}{2}$.

On the other hand, putting $\vartheta = 0$, we can solve equation
(\ref{ExpansionEquation}) explicitly for the radius $t$ as a function
of $r$:
\begin{equation}
  \label{0ExpansionEquation}
  \cosh 2t = \frac{4r^4 - 3r^2 + 1}{(1 - 4r^2)(1 - r^2)}
\end{equation}
This has real solutions for $0 < r < \frac{1}{2}$, and as $r$
approaches the upper limit $\frac{1}{2}$, $t$ becomes infinite. Thus,
for each value of $r$ between $0$ and $\frac{1}{2}$, there exists a
marginally trapped surface of radius $t$ given by this equation.

The two last paragraphs show that closed trapped surfaces of constant
radius exist for all $r < \frac{1}{2}$, but not for any larger $r$.
Now we will argue that this is true, not only for these $\psi$ and
$\varphi$ symmetric surfaces, but for all closed trapped surfaces with
respect to the $r$-foliation.  Namely, suppose that there exists a
non-symmetric closed surface ${\cal T}$ for $r > \frac{1}{2}$ whose
divergence $\vartheta_{\cal T} \leq 0$. The radius $t$ varies along
${\cal T}$ as a function of $\psi$ and $\varphi$, $t =
t(\psi,\varphi)$. This function reaches its maximum value $t_{max}$
for some angles $\psi_0$ and $\varphi_0$, $t_{max} = t(\psi_0,
\varphi_0)$. Now, consider the {\it symmetrical} closed surface whose
constant radius equals $t_{max}$. Clearly, this surface encloses
${\cal T}$, touching it in $(\psi_0, \varphi_0)$.  Since it by
construction lies outside ${\cal T}$, its outward orthogonal lightrays
in $(\psi_0, \varphi_0)$ must converge at least as fast as the ones
corresponding to ${\cal T}$. Hence, its divergence $\vartheta \leq
\vartheta_{\cal T} \leq 0$, and we would have found it in the analysis
above concerning such symmetrical surfaces. But we did not, and we may
conclude that there do not exist any closed trapped surfaces,
symmetrical or non-symmetrical, for $r > \frac{1}{2}$.

\begin{figure}[h]
\hfil
\begin{minipage}[b]{60mm} 
\caption{A Penrose diagram of the 3+1 dimensional growing black hole. Each
point represents a torus.}
\end{minipage}
\hfil
\begin{minipage}[b]{60mm}
  \vspace{20mm}
  \epsfbox{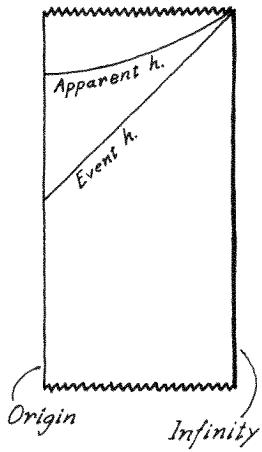}
\end{minipage}
\hfil
\end{figure}

We have then shown that the apparent horizon with respect to the
r-foliation lies at $r = \frac{1}{2}$. In the Penrose-diagram in
figure~10 we have drawn it together with the singularity and the
event horizon. The $\psi$ and $\varphi$ directions are suppressed
in this diagram, which means that each point actually represents a
torus.

\section{Some concluding remarks}
In this paper we have classified and discussed the possible causal
structures that can be obtained by identifying points in adS-space
connected by one of its symmetries. The method we have used rests on
the fact that the resulting spacetime may be written as a quotient
space $[adS]/{\cal G}_{_\Gamma}$ and therefore clearly should not
depend on which points are considered as identified with which---only
the symmetry $\Gamma$ matters while the choice of fundamental region
is irrelevant. In particular, as we saw in section~2, in order to
understand the causal behaviour it is sufficient to find the
hypersurface in adS-space where the Killing field corresponding to
$\Gamma$ is null.

Using these ideas we have generalized the different 2+1 dimensional
BTZ black holes to 3+1 dimensions. While the causal properties of the
two types of extremal cases---$M=0$ and $M=J$ in the notation
of Banados {\it et al} \cite{BH}---remains essentially unchanged as one extra
dimension is added, the non-extremal ones change dramatically: The
rotating black hole simply disappears, its event horizon being
destroyed by the extra dimension, while the non-rotating one becomes
non-static!

The 3+1 version of the non-rotating BTZ hole actually describes
a growing black hole with a toroidal event horizon. Furthermore. as
shown in the previous section, it possesses a non-trivial apparent
horizon---non-trivial in the sense that it does not coincide with the
event horizon, as is the case for example in the ordinary BTZ hole.
Thus, it has almost all characteristics of a realistic black hole
formed from gravitational collapse. Perhaps its most unphysical
property is the torus topology of infinity. On the other hand, the
other properties are largely due to precisely this toroidal nature.
For example, the closed trapped surfaces found above could not be both
closed and trapped was it not for their torus topology. This is because
in adS-space all trapped surfaces are non-closed (either $\bf R^2$ or
$\bf R^1 \times S^1$) and by means of identifications one could close
these into toruses, but never---as would be more realistic for a
trapped surface---into spheres.

Due to the, at the same time, simple and rich nature of this black
hole---simple because of its metric being that of adS-space, and rich
in the sense of describing a growing black hole with an apparent
horizon---it may turn out to be valuable as a toy model in different
areas. For example, it could play an important role in research concerning
quantum field theory and black hole physics.

But we also think that our methods are of some interest in their own right.
Usually one deduces the causal structure of a spacetime by studying
its metric. Then one has to worry about a number of coordinate related
problems, such as extendability of the spacetime or whether a
singularity is real or only signals a breakdown of coordinates.
However, as we have shown here, for spacetimes of the form $[{\cal M},
g^{\mu\nu}]/{\cal G}_{_\Gamma}$ (where the properties of $[{\cal M}, g^{\mu\nu}]$
are assumed to be known) it is not necessary to go through this type
of considerations---one does not even have to write down an explicit
metric.

However, our discussion has only concerned taking the quotient with
groups ${\cal G}_{_\Gamma}$ generated by a single generator. What if ${\cal
  G}_{_\Gamma}$ is generated by two or more generators? Are the same
methods still applicable? First, if ${\cal G}_{_\Gamma}$ is Abelian,
generated by $\Gamma_1$, $\Gamma_2$, etc., then $[{\cal M},
g^{\mu\nu}]/{\cal G}_{_\Gamma}$ could be equivalently constructed as
the quotient by one generator at a time, that is, as $(([{\cal M},
g^{\mu\nu}]/{\cal G}_{{_\Gamma}_1})/{\cal G}_{{_\Gamma}_2})$ etc., because,
since the generators commute, $\Gamma_2$ will be a symmetry also of $([{\cal M},
g^{\mu\nu}]/{\cal G}_{{_\Gamma}_1})$. Our method could then be
applied to one generator at a time, and thus extended to the case with
two or more commuting identifications.

On the other hand, for non-Abelian groups ${\cal G}_{_\Gamma}$ this
``quotient decomposition'' does not work because if $\Gamma_1$ and
$\Gamma_2$ are non-commuting, then $\Gamma_2$ is {\it not} a symmetry
of $[{\cal M},g^{\mu\nu}]/{\cal G}_{{_\Gamma}_1}$ (unless in
exceptional cases, in which the argument of the previous paragraph
goes through as before). Still, of course, $[{\cal M},
g^{\mu\nu}]/{\cal G}_{_\Gamma}$ is well-defined, but the
problem is that in the generic case it would not be clear how to
define the singularity. The view adopted in this paper, that the
singularity is the hypersurface beyond which the identification
Killing field $\xi$ is timelike does not makes sense any more when
there are more than one such field.

The reader may wonder why not the boundary to the region where at
least one of the generators in ${\cal G}_{_\Gamma}$ are timelike would
do as a definition for the singularity. Even if the consequences of
such a definition surely could be analysed, remember that part of the
motivation for defining the singularity as the boundary to the region
with timelike (identification) Killing flow, was that every closed
timelike curve in the quotient space actually must close {\it behind}
that surface, and that the exclusion of this ``timelike'' region
therefore guarantees the absence of closed timelike curves. But when
several identifications are performed, then, even if each of the
corresponding Killing fields are purely spacelike, the resulting
spacetime may contain closed timelike curves anyway. (A striking
example of this is the ``Gott time machine'', see ref. \cite{Gott}.)
For this reason, the motivation for our definition of the singularity
partly disappears for two or more non-commuting identifications.

Another possible definition for such situations---more in accord
with the one adopted in this paper---is the boundary to the smallest
region (in some sense) that one would have to remove from spacetime in
order to get rid of all closed timelike curves.  But it is difficult
to see how the concept of ``smallest region'' could be made precise,
and even if it could, it is not clear whether such a definition would
give a unique singularity.

We must conclude that it seems hard to construct any sensible
definition when the generators are non-commuting, except in special
cases. One such special case is when the identification Killing fields
have fixpoints, which then give rise to singularities in the quotient
space. Specific examples of this are provided by the multi black holes
discussed by Brill \cite{MultiBrill} and Steif \cite{MultiSteif}, or
the related black wormholes studied in ref. \cite{ViOchBrill}. Both of
these constructions make use of two or more of the static BTZ-type
Killing fields: Type $I_b$ when $a_1$ or $a_2$ equals zero. However,
because of what we have said here, we do not think it is possible to
generalize these spacetimes to rotating multi black holes or wormholes
with rotation, simply because the involved Killing fields would then
not have any fixpoints (except at infinity) and it would not any
longer be clear as to what the singularity is.

\section*{Acknowledgments}
We would like to thank Ingemar Bengtsson for many valuable discussions
and for his general support during the work with this paper. We are
also grateful for his, and Stefan {\AA}minneborg's, remarks and
suggestions concerning the manuscript.

\end{document}